\documentclass[pre,twocolumn,showpacs]{revtex4}

\usepackage[dvips]{graphicx}

\begin{document}

\title{Intelligent systems in the context of surrounding environment}
\date{5 April 2001; revised 29 June 2001}
\author{Joseph Wakeling}
\email{jwakeling@webdrake.net}
\affiliation{Department of Mathematics, Imperial College, 180 Queens Gate, London, SW7 2BZ, UK.}
\author{Per Bak}
\affiliation{Department of Mathematics, Imperial College, 180 Queens Gate, London, SW7 2BZ, UK.}

\begin{abstract}
We investigate the behavioral patterns of a population of agents, each controlled by a simple biologically motivated neural network model, when they are set in competition against each other in the Minority Model of Challet and Zhang.  We explore the effects of changing agent characteristics, demonstrating that crowding behavior takes place among agents of similar memory, and show how this allows unique `rogue' agents with higher memory values to take advantage of a majority population.  We also show that agents' analytic capability is largely determined by the size of the intermediary layer of neurons.

In the context of these results, we discuss the general nature of natural and artificial intelligence systems, and suggest intelligence only exists in the context of the surrounding environment (\emph{embodiment}).

Source code for the programs used can be found at \textbf{http://neuro.webdrake.net/}.
\end{abstract}


\maketitle

\section{Introduction}

Much research has been done into the computational possibilities of neural networks.  Yet the engineering and industrial applications of these models have often eclipsed their use in trying to come to an understanding of naturally occurring neural systems.

Whereas in engineering we often use single neural networks to attack a single problem, in nature we see neural systems in competition.  Humans, for example, invest in the stock market, attempt to beat their business rivals, or, in extreme examples, plan wars against each other.  We are, as Darwin identified a century and a half ago, in competition for natural resources; our neural systems---i.e.~our brains---are among the main tools we have to help us succeed in that competition.

In collaboration with Chialvo, one of the authors of this paper has developed a neural network model that provides a biologically plausible learning system~\cite{BC}, based essentially around `Darwinian selection' of successful behavioral patterns.  This simple `minibrain'---as we will refer to it from now on---has been shown to be an effective learning system, being able to solve such problems as the exclusive-\textsc{or} (\textsc{xor}) problem and the parity problem.  Crucially, it has also been shown to be easily able to \emph{un-learn} patterns of behavior once they become unproductive---an extremely important aspect of animal learning---while still being able to remember previously successful responses, in case they should prove useful in the future.  These capabilities, combined with the simplicity of the model, provide a powerful case for biological feasibility.

In choosing a competitive framework for this neural network, we follow the example of Metzler, Kinzel and Kanter~\cite{MKK}, using the delightfully simple model of competition within a population provided by the Minority Model of Challet and Zhang~\cite{CZ} (itself based on the `El Farol' bar problem created by Arthur~\cite{Art}).  In this game, a population of agents has to decide, independently of each other, which of two groups they wish to join.  Whoever is on the minority side `wins' and is given a point.  By combining these two models---replacing the fixed strategies of agents in Challet and Zhang's model with agents controlled by the Minibrain neural system---we have a model of neural systems in competition in the real world.

This is not the first model of coevolution of strategies in a competitive game---a particularly interesting example is Lindgren and Nordahl's investigation of the Prisoner's Dilemma, where players on a cellular grid evolve and mutate strategies according to a genetic algorithm~\cite{LN}.  However, we believe that the biological inspiration for the Minibrain model, and its demonstrated capacity for fast adaption, makes our model of special interest.

The structure of this paper is as follows: we begin with a discussion of what we mean when we talk about `intelligence', noting how historical influences have shaped our instinctive ideas on this subject in potentially misleading ways; in particular, we take issue with the suggestion that a creature's intelligence can be thought of as separate from its physical nature.  We suggest that intelligence can only be measured in the context of the surrounding environment of the organism being studied: we must always consider the \emph{embodiment} of any intelligent system.

This is followed by the account of the computer experiments we have conducted, in which we investigate the behavioral patterns produced in the Minibrain/Minority Model combination, and the ways in which they are affected by changing agent characteristics.  We show how significant crowding behavior occurs within groups of agents with the same memory value, and demonstrate how this can allow a minority of high-memory agents to take advantage of the majority population and `win' on a regular basis---and, by the same token, condemn a population of largely similar agents to continually losing.  Indeed, perhaps the most startling implication of this model is that, in a competitive situation, having a `strategy' might well prove worse than simply making random decisions.

These results are in strong contrast with those of Metzler, Kinzel and Kanter, whose paper inspired these experiments.  In their simulations, a homogeneous population of perceptron agents relaxes to a stable state where all agents have an average 50\% success rate, and overall population performance is better than random~\cite{MKK}.  The perceptrons learn, in effect, to produce an efficient market system, and do not suffer from the crowding effect produced by minibrain agents.  By the same token, however, it seems unlikely that a superior perceptron could win on a similar scale to a superior Minibrain.

We conclude with further discussion of the nature of intelligence, suggesting a conceptual approach that we believe will enable easier investigation of both natural and artificially created intelligent systems.  Having already suggested that we must consider `embodied' intelligences, we provide criteria for cataloguing that embodiment, consisting of hardwired parts---the input and output systems of the organism, the feedback mechanism that judges the success or failure of behavioral patterns---alongside a dynamic decision-making system that maps input to output and updates its methodology according to the signals received from the feedback system.

\section{What is `Intelligence'?}

The \emph{E.~Coli} bacterium has a curious mode of behavior.  If it senses glucose in its immediate surroundings, it will move in the direction of this sweet nourishment.  If it does not, it will flip over and move a certain distance in a random direction, before taking stock again, and so on and so on until it finds food.

Bacteria are generally not considered to be `intelligent'.  Yet this is a systematic response to environmental stimuli, not necessarily the \emph{best} response but nevertheless a \emph{working} response, a `satisficing' response.  The \emph{E.~Coli} bacterium is responding in an intelligent way to the problem of how to find food.  How do we square this with our instinctive feeling that bacteria are \emph{not} intelligent?  Are our instincts mistaken?  How, instinctively, do we define intelligence?

Historically, philosophers have often proposed the idea of a separation between `body' and `mind'.  The human mind, from this point of view, is something special, something distinct, something not bound up in the messy business of the real world.  It's this, we are told, that separates us from the animals: we have this magical ability to \emph{understand}, to \emph{think}, to \emph{comprehend}---the ability to view the world in a rational, abstract way and thus arrive at some fundamental \emph{truth} about how the universe works.

The idea of separate compartments of reality for body and mind has lost its stranglehold over our way of thinking, but its influence lingers on in our concept of intelligence.  Our minds, our consciousness, may be the result of physical processes, but we still cling to the idea that we have the ability to discover an abstract reality, and it's this idea that informs our notion of `intelligence'.  An intelligent being is one that can see beyond its own personal circumstances, one that is capable of looking at the world around it in an objective fashion.  Given enough time, it can (theoretically) solve any problem you care to put before it.  It is capable of rising above the environment in which it exists, and comprehending the nature of True Reality.

Naturally, this has informed our ideas about artificial intelligence.  An artificially intelligent machine will be one that works in this same environmentally uninhibited manner.  If we tell it to drive a car, it will be able (given time to teach itself) to drive a car.  If we tell it to cook a meal, it will be able to cook a meal.  If we tell it to prove Fermat's Last Theorem~\ldots\  All of these, of course, assume that we have given it some kind of hardware with which to gather input and make output to the relevant system, whether car, kitchen or math textbook---assume, indeed, that we have these systems present at all---and it's this necessity that causes us to realize that in fact, \emph{the mind and its surrounding environment (including the physical body of the individual) are inseparable.}  Our brains are the product of evolution; they are not an abstract, infinite system for solving abstract, infinite problems, but rather a very particular system for solving the very particular problems involved in coping with the environmental pressures about us.  In this respect, we're no different from the \emph{E.~Coli} bacterium we discussed earlier: the environments we inhabit are different, and consequently so are our behavioral patterns, but on a conceptual level there is nothing to choose between us.

\emph{Intelligence only exists in the context of its surrounding environment.}  So, if we are to attempt to create an artificial intelligence system, we must necessarily also define a world in which it will operate.  And the question of \emph{how} intelligent that system is can only be answered by examining how good it is at coping with the problems this world throws up, by its ability to utilize the data available to it to find working solutions to these problems.

\section{`Minibrain' Agents in the Minority Model}

The `minibrain' neural system, developed by one of the authors in collaboration with Chialvo~\cite{BC}, is an extremal dynamics-based decision-making system that responds to input by choosing from a finite set of outputs, the choice being determined by Darwinian selection of good (i.e.~successful) responses to previous inputs (negative feedback).  We use the simple layered version of this model, consisting of a layer of input neurons, a single intermediary layer of neurons, and a layer of output neurons; each input neuron is connected to every intermediary neuron by a synapse, and similarly each intermediary neuron is connected to every output neuron.  Every synapse is assigned a `strength', initially a random number between $0$ and $1$.

Competing against each other in the Minority Model, each agent receives data about the past, and gives as output which of the two the groups---we label them $0$ and $1$---that it wishes to join.  We follow the convention of Challet and Zhang's version of the game, that this knowledge is limited to knowing which side was the minority (i.e.~winning) group at each turn in a finite number of past turns~\cite{CZ}, so that agent input can be represented by a binary number of $m$ bits, where $m$ is the agent's memory.  So, for example, if in the last three turns group $0$ lost, then won, then won again, this would be represented by the binary number $110$, where the left-most bit represents the most recent turn, and each bit is determined by the number of the \emph{losing} (majority) group that turn (we choose these settings in order to match the way our computer code is set up).

\begin{figure}[t]\includegraphics[width=84.0mm, height=44.2mm]{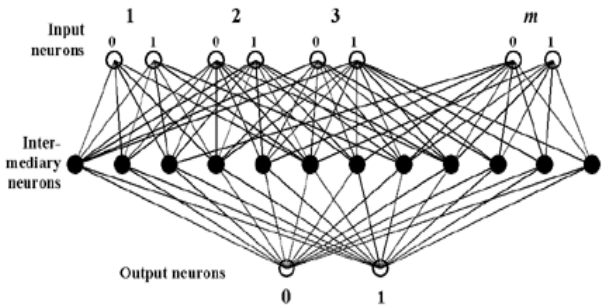}\caption{\label{fig:1}~Architecture of minibrain agents.  Every input neuron is connected to every intermediary neuron, and every intermediary neuron is connected to every output neuron.  For our setup, we have two outputs, and $2m$ inputs, where $m$ is the agent's memory.}\end{figure}

In order to preserve symmetry of choice between the two groups, an agent with a memory of $m$ turns will have $2m$ input neurons, with the first of the $i$th pair of neurons firing if the bit representing the result $i$ turns ago is $0$, the second neuron of the $i$th pair firing if the result was $1$.  For example, if an agent with a memory of $3$ (and hence with $6$ input neurons) is given the past $110$ as we discussed above, then the second, fourth and fifth input neurons will fire.  Fig.~\ref{fig:1} gives a picture of this architecture (to avoid over-complicating the diagram, not all connections are shown).

To determine the intermediary neuron that fires, we take for each the sum of the strengths of the synapses connecting it to the firing input neurons.  The intermediary neuron with the greatest such sum is the one that fires.  Then, the output neuron that fires ($0$ or $1$) is the one connected to the firing intermediary neuron by the strongest synapse.

Each turn, the synapses used are `tagged' with a chemical trace.  If the output produced that turn is satisfactory (in this setup, if the agent joins the minority group), no further action is taken.  If the output is not satisfactory, however, a global feedback signal (e.g.~the release of some hormone) is sent to the system, and the tagged synapses are `punished' for their involvement in this bad decision by having their strengths reduced (in our model, by a random number between $0$ and $1$).  As we noted in the Introduction, this Darwinian selection of successful behavioral patterns has already been shown to be an effective learning system when `going solo'~\cite{BC}; how will it cope when placed in competition?

\begin{figure}[t]\includegraphics[width=75.4mm, height=51.9mm]{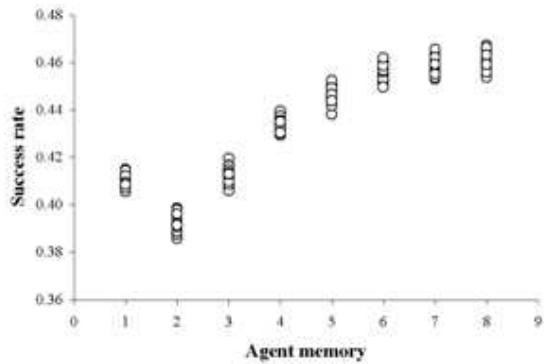}\caption{\label{fig:2}~Success rates of a mixed population of minibrain agents against their memory.  Agents have $48$ intermediary neurons.}\end{figure}

Fig.~\ref{fig:2} shows the success rates of agents of different memory values.  A group of $251$ agents has an even spread of memory values between $1$ and $8$; each agent has $48$ intermediary neurons.  The figure shows their success rates over a period of $2\times 10^{4}$ turns.

To a certain extent, these results reflect those found by Challet and Zhang when they explore the behavior of a mixed population of fixed-strategy agents~\cite{CZ}, inasmuch as performance improves with higher memory but tends to saturate eventually.  Standard deviation within each memory group is much lower for minibrain agents, however, suggesting crowding behaviour within memory groups, and we will later show that this does indeed occur.

Disappointingly, we see that not one agent achieves as much as a 50\% success rate---they would all be better off tossing coins to make their decisions.  The even spread of memory values throughout the population means that agents with higher memory values cannot take full advantage of their extra knowledge: the crowding behavior between agents with the same memory cancels out most of the positive effects.  It's no good having lots of data on which to base your decision if lots of other people have that same data---everyone will come to the same conclusion and, in the Minority Model, that means losing.

\begin{figure}[t]\includegraphics[width=75.4mm, height=60.0mm]{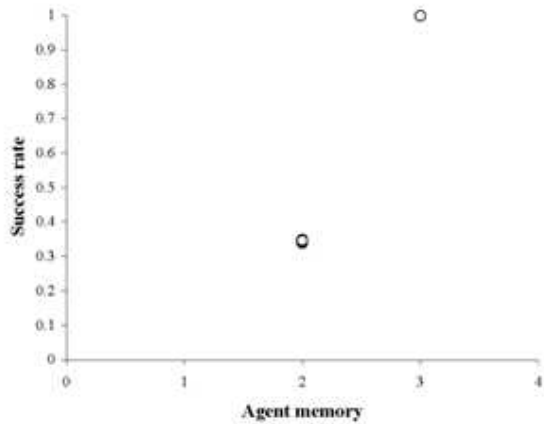}\caption{\label{fig:3}~Success rate of a single agent of memory $3$, versus a $250$-strong population of memory $2$.  Agents have $48$ intermediary neurons.}\end{figure}

Necessarily, then, one of the conditions for an agent to succeed---i.e.~to beat the coin-tossing strategy---is that there must be few other agents with the same amount of memory.  This is demonstrated starkly in Fig.~\ref{fig:3}, displaying the results for a population of $251$ agents of whom \emph{one} has a memory of $3$, the rest only $2$.

The astonishing success of this `rogue' agent (it makes the right decision approximately 99.8\% of the time) shows clearly just how important a factor this crowding behavior is in the success (or failure) of agents.  The fact that this agent is the only one receiving the extra data means that he can use it to his advantage.  Contrast this with the other agents who, for all their careful thinking, fail miserably because \emph{almost all of them think alike---entirely independently---almost all of the time.}

\begin{figure}[t]\includegraphics[width=75.4mm, height=51.2mm]{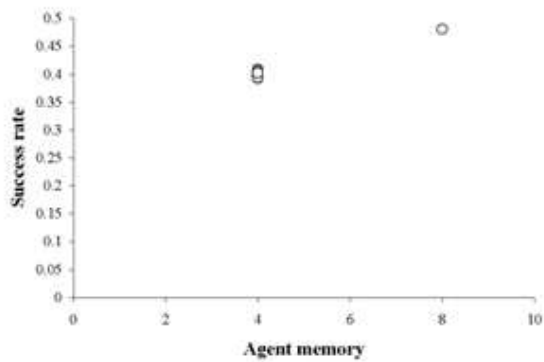}\caption{\label{fig:4}~Success rate of a single agent of memory $8$, versus a $250$-strong population of memory $4$.  Agents have $48$ intermediary neurons.}\end{figure}

This example leads us to ask the more general question: given a population of agents who all have memory $m$, can we always find such a `rogue', an agent capable of understanding and thus beating the system?  That it is not merely a matter of agent memory is amply demonstrated in Fig.~\ref{fig:4}, where we see a population of memory $4$ pitted against a rogue with a memory of $8$.

Despite its high memory value (twice that of the majority population) the rogue agent is unable to beat the coin-tossing strategy.  Why is this?  A higher memory value should, by our earlier results, always be an advantage.  Certainly, since we have respected symmetry of choice between agent outputs, it should not be a \emph{disadvantage}.  What factor is it that prevents this agent from making full use of the memory available to it, memory which surely has within it useful data patterns that predict the behavior of the agents with memory $4$, and thus should allow the rogue agent the success we expect it to achieve?

The answer becomes clear when we examine the nature of the input that each agent receives---a binary number of length $m$, where $m$ is the value of the agent's memory.  So, it follows that the total possible number of inputs will be $2^{m}$.  For an agent with memory $4$, this means $16$ possible inputs.  For an agent with memory $8$, the total number of possible inputs is $256$.  Compare this to the number of \emph{intermediary} neurons possessed by each agent ($48$, in all the simulations we've run so far) and we realize that, while this is an adequate number for an agent receiving $16$ different possible inputs, it is wholly inadequate for an agent having to deal with some $256$ possible inputs.  The number of intermediary neurons restricts the maximum performance of an agent by placing a limit on the amount of memory that can be effectively used.

\begin{figure}[t]\includegraphics[width=75.4mm, height=65.0mm]{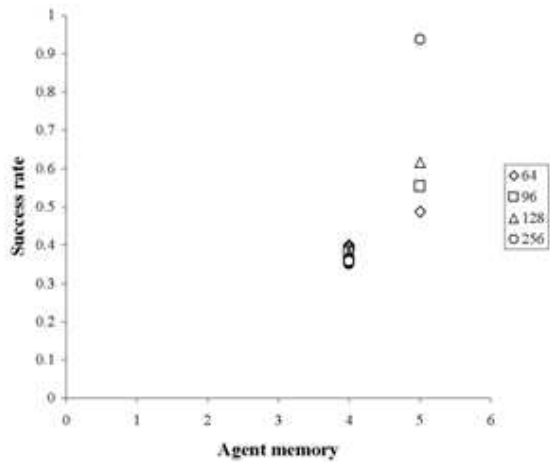}\caption{\label{fig:5}~Success rates of single agents of memory $5$, versus a $250$-strong population of memory $4$, in simulations with $64$, $96$, $128$ and $256$ intermediary neurons per agent.}\end{figure}

Bearing this condition in mind, we run a new set of games, again with a majority population of memory $4$, but this time with a rogue of memory $5$, and with the number of intermediary neurons given to each agent varying in each of these games.  Fig.~\ref{fig:5} shows the results of games where agents have intermediary layers of, respectively, $64$, $96$, $128$ and $256$ neurons.

The implications are clear---it is the number of intermediary neurons, as well as the amount of memory, that control whether or not a rogue agent can succeed, and, if it does, by how much.  A higher memory value will always be an advantage, but the degree to which it is advantageous will be determined by the number of intermediary neurons possessed.  Memory, obviously, determines how much information an agent can receive; the intermediary neurons are what provide agents' analytic capability.

Our computer simulations suggest that in situations such as the ones already discussed, with a majority population of memory $m$, it is the intermediary neurons, rather than the amount of memory possessed, that control the ability of a rogue agent to succeed.  A memory of $m+1$ is all that is required, \emph{provided} the rogue has enough intermediary neurons to be able to use it effectively.

We can muddy the waters, so to speak, by giving the majority population an evenly distributed \emph{spread} of memory values (perhaps from $1$ to $m$) rather than a single value.  Where a single memory value is used, the crowding behavior observed within memory groups will easily allow rogue agents to predict the minority group.  With a series of different, smaller groups in competition, it becomes significantly less easy to make accurate predictions, and rogue agent success rates fall significantly.  Herding sheep is fairly easy; jumping into the middle of a brawl is dangerous for anyone, even a world champion martial artist.

All things considered, it seems as though this may be the key point in determining agent success.  An agent can only be truly successful if it has plenty of `prey' whose weaknesses it can exploit.  If the behavior of the prey is highly unpredictable, or the prey are capable of biting back, the agent's chances of success are vastly reduced.

\section{Analysis of crowding behavior within memory groups}

We have on several occasions referred to crowding behavior of minibrain agents within the same memory group.  In this section, we give a brief mathematical analysis of what causes this to arise.

We begin with a simple case, assuming that all agents have the same memory value.  Obviously, because of the nature of the game, a majority of these agents will behave in the same way each turn.  What we show, however, is that this majority is significantly more than would be found if the agents were deciding randomly as to which group to join.  Were agents to employ this strategy, the mean size of the (losing) majority group would be only a little over 50\%.

We define by $0 \leq x_{i} (I) < 0.5$ the proportion of agents in the minority group given input $I$, where the subscript $i$ is the number of times input $I$ has occurred before.  If an input has not been seen before by agents, it follows that they will decide randomly which group to join, and so we have $x_{0} (I) \simeq 0.5$ for all possible inputs $I$.

If an input \emph{has} been seen before, it follows that those agents in the minority group on that occasion --- i.e.~those who were successful---will make the same decision as last time.  Those who were unsuccessful last time will make a random decision as to which group to join: we can expect, on average, half of them to change their minds, half to stay with their previous choice.

The effect of this, ironically, is that this last group---the unsuccessful agents who keep with their previous choice---will probably (in fact, almost certainly) form the minority group this time round.  And so we can define a recurrence relation,
\begin{displaymath}\qquad x_{i+1} (I)\simeq\frac{1}{2}\left(1 - x_{i} (I)\right)\end{displaymath}

determining the expected proportion of agents joining the minority group for each occurrence of input $I$.  This allows us to develop a more general equation,
\begin{displaymath}\qquad x_{i+1} (I)\simeq\varphi (i, I)\end{displaymath}

where
\begin{displaymath}\qquad\varphi (i,I) = \frac{1}{3} \left(\frac{2^{i} + (-1)^{i-1}}{2^{i}}\right) + \frac{(-1)^{i}}{2^{i+1}} \left( 1-x_{0} (I)\right)\end{displaymath}

Observe that this holds for $i=0$, as a little calculation reveals $x_{1} (I)\simeq\frac{1}{2}\left( 1-x_{0} (I)\right) =\varphi (0,I)$.  Now, assume the equation holds for $i = n-1$, with $n$ any positive integer, so $x_{n} (I)\simeq\varphi (n-1,I)$.

By the recurrence relation,
\begin{displaymath}\qquad x_{n+1} (I) \simeq\frac{1}{2}\left( 1-x_{n} (I)\right) =\frac{1}{2} \left( 1-\varphi (n-1,I)\right)\end{displaymath}

\begin{displaymath}\qquad =\frac{1}{2}\left[ 1-\left\{\frac{1}{3} \left(\frac{\begin{scriptstyle}2^{n-1} + (-1)^{n-2}\end{scriptstyle}}{2^{n-1}}\right) + \frac{\begin{scriptstyle}(-1)^{n-1}\end{scriptstyle}}{2^{n}} \left(\begin{scriptstyle} 1-x_{0} (I)\end{scriptstyle}\right)\right\}\right]\end{displaymath}

\begin{displaymath}\qquad =\frac{1}{2}\left[\frac{1}{3} \left(\frac{\begin{scriptstyle} 3\times 2^{n-1} -2^{n-1} +(-1)^{n-1}\end{scriptstyle}}{2^{n-1}}\right) + \frac{(-1)^{n}}{2^{n}} \left(\begin{scriptstyle} 1-x_{0} (I)\end{scriptstyle}\right)\right]\end{displaymath}

\begin{displaymath}\qquad =\frac{1}{2}\left[\frac{1}{3} \left(\frac{2^{n} +(-1)^{n-1}}{2^{n-1}} \right) + \frac{(-1)^{n}}{2^{n}} \left( 1-x_{0} (I)\right)\right]\end{displaymath}

\begin{displaymath}\qquad =\frac{1}{3} \left(\frac{2^{n} +(-1)^{n-1}}{2^{n}} \right) + \frac{(-1)^{n}}{2^{n+1}} \left( 1-x_{0} (I)\right)\end{displaymath}

\begin{displaymath}\qquad =\varphi (n,I)\end{displaymath}

Hence, $x_{n+1} (I)\simeq\varphi (n,I)$, and so by the induction hypothesis $x_{i+1} (I)\simeq\varphi (i,I)$  for all $i\geq 0$.

It follows, then, that as $i\to\infty$, so $x_{i} (I)\to\frac{1}{3}$, and so, with repeated exposure to the input $I$, we will find that on average $\frac{2}{3}$ of the agents will produce the same output.  As a result, the average majority size per turn (regardless of input given) will also tend to $\frac{2}{3}$ as the agents become saturated by all the possible inputs.

\begin{figure}[t]\includegraphics[width=75.4mm, height=51.4mm]{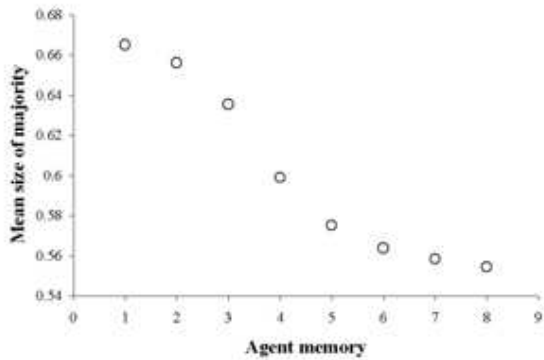}\caption{\label{fig:6}~Mean size of majority each turn in games with uniform agent memory, against different choices for this memory value.  Agent population per game is $251$, but majority size is given proportionally.  Agents have $48$ intermediary neurons.}\end{figure}

This can be observed in Fig.~\ref{fig:6}, which shows the average proportion of agents joining the majority group each turn in eight different games involving single memory value populations, the first involving agents of memory $1$, the second with agents of memory $2$, and so on up to the final game, with agents of memory $8$.  Each game takes place over a time period of some $5\times 10^{3}$ turns.

As memory increases, so the number of possible inputs also increases, meaning there is less repeated exposure to individual inputs, and hence less crowding for a given time period.  Within a time-scale of $5\times 10^{3}$ turns, the behavior of agents with longer memories is random more often than not, and so the mean size of majority is similar to that of agents making random decisions.  As the number of turns increases, so we can expect the mean size of the majority to tend to $\frac{2}{3}$ for all memory values, not just the lowest.

What implications does this have for games involving a mixed population of agents, such as that displayed in Fig.~\ref{fig:2}?  Overall, the same principles will apply.  Repeated exposure to the same input will produce the same crowding effect.  But we note that the inputs given to this system---eight-digit binary numbers---are interpreted differently by different agents.  For agents with lower memories, many of these `different' inputs are interpreted as being the same!  For example, the inputs $11010010$ and $11001011$ are the same to an agent with a memory of $3$ or less.  So---as is demonstrated by Fig.~\ref{fig:6}---the crowding effect surfaces earlier in agents with lower memory values, and hence they are adversely affected to a greater degree.

The agents with higher memory values fail to beat the 50\% success rate, however, because there are too many of them---any insights they might have into the crowding behavior of the lower memory groups are obscured by the actions of their fellow high-memory agents.  Thus, the kind of behavior we see in Fig.~\ref{fig:2}: the lower memory agents perform the worst, with the success rate increasing towards some `glass ceiling' as agent memory increases.  It's only unique `rogue' agents, who don't have a large group of fellows, who can see the crowding effect and thus beat the system.

Even such rogue agents cannot succeed by any great margin in the case where they are pitted against a spread of memory values.  The crowding behavior of the individual groups is obscured by the large number of them and predictions become difficult; the rogue has to work out, not just in which direction the crowding within each group will go, but how much crowding will be taking place in each group---a difficult task indeed!

\begin{figure}[t]\includegraphics[width=75.4mm, height=65.0mm]{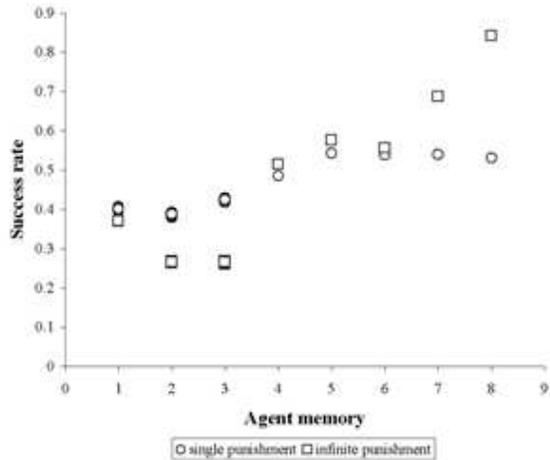}\caption{\label{fig:7}~Success rates of rogue agents of memory $5$--$8$, versus a majority population with memory $1$--$3$, in games involving single punishment and `infinite' punishment of unsuccessful synapses.  Total agent population is $251$.  Agents have $256$ intermediary neurons.}\end{figure}

If we increase the crowding, we also increase the rogue's chances of success.  Fig.~\ref{fig:7} shows the results from two different games involving $251$ agents.  Five of them are `rogue' agents with memory values of, respectively, $4$, $5$, $6$, $7$ and $8$.  The rest have an even spread of memory values from $1$ to $3$.  In order to allow the higher memory values to be useful, we give agents $256$ intermediary neurons.  The difference between the games is that in the first, when punishing unsuccessful synapses, we employ the principle that has been used throughout this paper---synapses are punished \emph{once}.  In the second game, the punishment does not stop until the agent has learned what would have been the correct output.  The result is that, when an input has been seen before, we will have 100\% agreement within memory groups.

We can see here how the increased crowding caused by `infinite' punishment allows the rogues to take advantage and be successful.  A higher memory value is required for substantial success, but substantial success is possible---at the expense of the lower memory groups, whose success rates are substantially decreased by the extra crowding behavior they are forced to produce.  The rogue agents in the game with single punishment, by contrast, are barely able to do better than a 50\% success rate---though they can evidently glean \emph{some} data from the crowding behavior displayed by the lower memory groups, it's not sufficient for any great success and they are only barely able to beat the expected success rate were they to make purely random decisions.

This is a striking result, to say the least.  \emph{The inevitable consequence of an analytic strategy is a predisposition to failure.}  Challet and Zhang~\cite{CZ} and W. B. Arthur~\cite{Art} have already shown that fixed strategies can prove to be a disadvantage compared to random decisions; this occurs when the number of available strategies is small compared to the number of agents.  The crowding behavior that results from minibrain agents' imperfect analysis will inevitably reduce the number of strategies in use, thus dooming themselves to worse-than-random results.

We can see this at work in the real world, every day.  Many strategies---whether for investments, business strategies, forming a relationship, or any of the myriad problems we have to solve---fail, because they are based on common knowledge, and as such, will be similar to most other people's strategies.  We are often told, `Everybody knows that \ldots ', but few people realize the negative side of following such advice: since \emph{everybody} knows it, \emph{everybody} will come to the same conclusions as you, and so your strategy will be unlikely to succeed.  Perhaps the best recent example is the internet boom-and-bust: so many people thought the internet was the place to invest, the market overheated, and many companies went belly-up.

As this paper was being prepared, a report was broadcast on UK television about an experiment in which a four-year-old child, picking a share portfolio, was able to outdo highly experienced City traders on the stock market.  In such systems, with everyone's imperfect analysis competing against everyone else's, it seems highly likely that random decisions sometimes really are the best; the Minibrain/Minority Model combination would appear to confirm this.

\section{`Intelligence' Reconsidered}

Another interesting conclusion to be drawn from the computer experiments here described is that, given some particular Minibrain agent, there is no way of deciding if it will be successful or not unless we know about the other agents it will be competing with.

In a sense this is not surprising.  We know, for example, that to be a high-flier at an Ivy League university requires considerably more academic ability than most other educational institutions.  The athlete coming last in the final of the Olympic 100m can still run faster than almost anyone else in the world.  We know that \emph{in these contexts} the conditions to be the `best' are different, but there is surely still an overall comparison to be made between the whole of humanity.  Or is there?  Recall our suggestion in the introduction to this paper that the question of how intelligent a system is can only be answered by examining how good it is at coping with the problems its surrounding environment throws up.  To return to Minibrain agents: by the concepts we discussed earlier, it is agents' \emph{intelligence}, and not just their success rate, that is dependent on their fellows', as well as their own, characteristics; indeed, the two measures---success and intelligence---cannot be separated.

Contrast this with how we have identified a whole range of factors---memory, the number of intermediary neurons, the amount of punishment inflicted on unsuccessful synapses---that affect the manner in which an agent performs.  There are objective comparisons that can be made between agents.  While we might accept that any measure of `intelligence' we can conceive of will only hold in the context of the Minority Model, surely it is not fair to suggest that the only valid measure of intelligence is success rate in the context of the population of agents we place within that world?

Before we rush off to define our abstract `agent IQ', however, it's worth noting that all the measures of \emph{human}, as well as Minibrain, intelligence that we have put in place are in fact measures of success in particular contexts.  When a teacher calls a pupil a `stupid boy', he is not commenting on the child's intelligence in some abstract sense, but rather the child's ability to succeed at the tasks he is set in the school environment.  (Einstein was considered stupid in the context of a school environment where dyslexia and Asperger's syndrome were unknown.)  When we say that human beings are more intelligent than other animals what we in fact mean is that human beings are more successful at manipulating their environment to their own benefit.  High-fliers at Ivy League universities are considered intelligent because of their academic success.  Olympic athletes are considered intelligent in the context of their sport because they are capable of winning consistently.

Even human IQ tests, long thought to provide an abstract and objective measure of intelligence, work in this fashion, being a measure of an individual's success in solving logical problems.  More recently these tests have been shown to discriminate against certain individuals based on their cultural background---a further indication of their non-abstract, non-objective nature---and in addition to this, psychologists are now proposing that there are other forms of intelligence, for example \emph{emotional} intelligence or `EQ', which are just as important to individual success as intellectual ability.

Were abstract measures of intelligence possible, it would be reasonable to ask: `Who was more intelligent, Albert Einstein or Ernest Shackleton?'  As it turns out, this question is impossible to answer.  Shackleton probably lacked Einstein's capacity for scientific imagination, Einstein probably didn't know a great deal about arctic survival, but both were highly successful---and thus by implication intelligent---\emph{in the context of their own chosen way of life.}  The same is true of our hypothetical Ivy League student and Olympic runner.  We suggest that no other possible measure of intelligence is truly satisfactory.

It is not an entirely easy concept to take on board.  In particular, it conflicts with our instinctive sense of what it means to be `intelligent'.  Casually---and not so casually---we talk about people's intelligence in the context of their \emph{understanding}, their \emph{conceiving}, their \emph{awareness}.  In other words, we talk about it in the context of their \emph{consciousness}.  In their paper `Consciousness and Neuroscience'~\cite{CK}, Francis Crick and Cristoph Koch refer to the philosophical concept of a `zombie', a creature that looks and acts just like a human being but lacks conscious awareness.  Using the concepts of intelligence we have been discussing, this creature is just as intelligent as a real human.

Yet, on closer examination, this is not such an unreasonable idea.  Such a `zombie' is probably scientifically untenable, but it should be noted that our measures of `intelligence' do not measure consciousness, at least not explicitly.  A digital computer can solve logical problems, for example, and it seems very unlikely that such computers are conscious.  The `emotional intelligence' we referred to earlier almost certainly has some unconscious elements to it: our ability to respond to a situation in an appropriate emotional manner tends to be an instinctive, more than a conscious, response.  Lizards, it is thought, lack a conscious sense of vision but they can still catch prey, find a mate and so on, using their visual sense to do so.  In fact, most of the organisms that exist on Earth are probably not conscious.  Consciousness, most likely, is a product of brain activity that is a useful survival aid, a useful aid for success.  An \emph{aid} for success, and thus for intelligence, rather than a requirement.

How, then, should we approach the question of what is an intelligent system?  In their description of the construction of the Minibrain neural system, Bak and Chialvo note: `Biology has to provide a set of more or less randomly connected neurons, and a mechanism by which an output is deemed unsatisfactory \ldots .  It is absurd to speak of meaningful brain processes if the purpose is not defined in advance.  The brain cannot learn to define what is good and what is bad.  This must be given at the outset.  From there on, the brain is on its own'~\cite{BC}.  These concepts provide us with a way of thinking about intelligent systems in general, whether naturally occurring biological systems or man-made artificial intelligence systems.

An intelligent system might be thought of as consisting of the following parts:

(i)\emph{A hardwired set of inputs and outputs, which the system cannot change.}  It can perhaps change which of them it takes notice of and which of them it uses, but its options are fixed and finite.

(ii)\emph{A decision-making system.}  Given an input, a systematic process is applied to decide what output to make.  This can range from the purely deterministic (e.g.~a truth-table of required output for each given input) to the completely random.  The E Coli bacterium's behavior in response to the presence or otherwise of glucose---either moving in the direction of the food or, if none is to be found, in a random direction---is a perfect example.

(iii)\emph{A hardwired system for determining whether a given output has been successful, and sending appropriate feedback to the system.}  Again, the nature of this can vary.  In our computer experiments, success is defined as being in the minority group.  For the \emph{E.~Coli} bacterium, success is finding food.  Possible types of feedback range from the positive reinforcement of successful behavior practiced by many neural network systems, to the negative feedback of the Minibrain model.  The \emph{E.~Coli} bacterium provides perhaps the most extreme example: if it doesn't find food within a certain time period, it will die!

The last of these is perhaps the most difficult to come to terms with, simply because as human beings, we instinctively feel that it is a \emph{conscious} choice on our part as to whether many of our actions have been successful or not.  Nevertheless, the ultimate determination of success or failure must rest with hardwired processes over which the decision-making system has no control.  If nothing else, we are all subject to the same consideration as \emph{E.~Coli}: if our actions do not provide our physical bodies with what they need to survive, they, and our brains and minds with them, will perish.

We should, perhaps, include an extra criterion that for a system to be \emph{truly} intelligent, the feedback mechanism must in some way affect the operation of the decision-making system, whether it is punishing `bad' synapses in the Minibrain neural network, changing the entries in a truth-table, or killing a bacterium.  A system that keeps making the same decision regardless of how consistently successful that decision is, isn't being intelligent.  With this in mind, we might consider systems such as \emph{E.~Coli} (i.e.~systems which employ one single strategy, and when it becomes unsuccessful simply \emph{stop}) to be \emph{minimally intelligent} systems.  They're nowhere near as smart as other systems, natural and artificial, but at least they know when to quit.

Intelligence, we suggest, is not an abstract concept.  The question of what is intelligent behavior can only be answered in the context of a problem to be solved.  So in the search for artificial intelligence, we must necessarily start with the world in which we want that intelligence to operate; we cannot begin by creating some `consciousness-in-a-box' to which we then give a purpose, but must first establish what we want that intelligence to \emph{do}, before building the systems---input-output, decision-making, feedback---that will best achieve that aim.  Computer programmers already have an instinctive sense of this when they talk about, for example, the `AI' of a computer game.  (Purpose: to beat the human player.  No longer the deterministic strategies of Space Invaders---many modern computer games display a great subtlety and complexity of behavior.)  This is not to denigrate attempts to build conscious machines.  Such machines would almost certainly provide the most powerful forms of artificial intelligence.  But we are still a long way from understanding what consciousness is, let alone being able to replicate it, and as we have noted here, consciousness is not necessarily needed for intelligent behavior.

The experiments discussed in this paper involve `toy' models.  Comparing the Minibrain neural system to the real human brain is like comparing a paper airplane to a jumbo jet~\cite{BC}.  But paper airplanes can still fly, and there are still lessons to be learned.  These `toy' experiments provide us with a context to begin identifying what it means to be intelligent.  We have been able to suggest criteria for identifying intelligent systems that avoid the controversial issues of consciousness and understanding, and a method of determining how intelligent such systems are that rests on one simple, useful and practical question: how good is this system at doing what it's meant to do?  In other words, we and others have begun to demystify the subject of intelligence and maneuver it into a position where we can begin to ask precise and well-defined questions.  Paper airplanes can fly for miles if they're launched from the right places.

~\\
~

\end{document}